\documentclass[conference]{IEEEtran}
\IEEEoverridecommandlockouts
\usepackage{algorithm}
\usepackage{arevmath}     
\usepackage[noend]{algpseudocode}
\usepackage{mathptmx} 
\usepackage[dvipsnames,table,xcdraw]{xcolor}
\usepackage{amsmath}
\DeclareMathOperator*{\argmax}{arg\;max}
\DeclareMathOperator*{\argmin}{arg\;min}
\usepackage{mathtools}
\usepackage{booktabs} 
\usepackage{ctable}
\usepackage{multirow}
\usepackage{multicol}
\usepackage{booktabs}
\usepackage{xspace}
\usepackage{graphicx}
\usepackage[caption=false]{subfig}
\usepackage{listings}
\usepackage{fancyhdr}
\usepackage[normalem]{ulem}
\usepackage[hyphens]{url}
\usepackage[sort,nocompress]{cite}
\usepackage[final]{microtype} 
\usepackage{flushend}
\usepackage{textgreek}
\usepackage{lipsum}

\usepackage{pifont}
\usepackage{glossaries}
\usepackage{acronym}
\usepackage[bookmarks=true,breaklinks=true,colorlinks,linkcolor=black,citecolor=blue,urlcolor=black]{hyperref}
\usepackage[capitalise]{cleveref} 
\usepackage[moderate]{savetrees}
\usepackage{threeparttable}
\def\BibTeX{{\rm B\kern-.05em{\sc i\kern-.025em b}\kern-.08em
    T\kern-.1667em\lower.7ex\hbox{E}\kern-.125emX}}
\newcommand\sanity[1]{{\color{black}\textbf{}#1}}

\newcommand{\eg}{\textit{e.g.,}}
\newcommand{\ie}{\textit{i.e.,}}
\newcommand{\relu}{\gls{relu}}
\newcommand{\gelu}{\gls{gelu}}
\newcommand{\silu}{\gls{silu}}
\newcommand{\prj}{\textit{Flex-SFU}}

\newacronym{ai}{AI}{artificial intelligence}
\newacronym{cpu}{CPU}{central processing unit}
\newacronym{tpu}{TPU}{tensor processing unit}
\newacronym{vpu}{VPU}{vector processing unit}
\newacronym{soa}{SoA}{state-of-the-art}
\newacronym{mxu}{MXU}{matrix-multiply unit}
\newacronym{pwl}{PWL}{piecewise linear}
\newacronym{dnn}{DNN}{deep neural network}
\newacronym{relu}{\texttt{ReLU}}{rectified linear unit}
\newacronym{gelu}{\texttt{GELU}}{Gaussian error linear unit}
\newacronym{silu}{\texttt{SiLU}}{sigmoid linear unit}
\newacronym{lut}{LUT}{lookup table}
\newacronym{madd}{MADD}{multiply-add}
\newacronym{msb}{MSB}{most significant bit}
\newacronym{simd}{SIMD}{single instruction multiple data}
\newacronym{mse}{MSE}{mean squared error}
\newacronym{mae}{MAE}{maximum absolute error}
\newacronym{aae}{AAE}{average absolute error}
\newacronym{ulp}{ULP}{unit of least precision}
\newacronym{rvv}{RVV}{RISC-V vector extension}
\newacronym{isa}{ISA}{instruction set architecture}
\newacronym{ram}{RAM}{random access memory}
\newacronym{bst}{BST}{binary search tree}
\newacronym{pnr}{PnR}{place-and-route}
\newacronym{nlp}{NLP}{natural language processing}
\newacronym{atc}{ATC}{Ascend Tensor Compiler}
\newacronym{rtl}{RTL}{register transfer level}
\newacronym{sfu}{SFU}{special function unit}
\newacronym{dcu}{DCU}{data control unit}
\newacronym{adu}{ADU}{address decoding unit}
\newacronym{ltc}{LTC}{lookup table cluster}

\begin{document}

\title{\prj: Accelerating DNN Activation Functions by Non-Uniform Piecewise Approximation
}

  \author{\IEEEauthorblockN{
      Enrico Reggiani\IEEEauthorrefmark{1}\IEEEauthorrefmark{2}\textsuperscript{1}, 
      Renzo Andri\IEEEauthorrefmark{1}\textsuperscript{1},
      Lukas Cavigelli\IEEEauthorrefmark{1}}
      \IEEEauthorblockA{\\\IEEEauthorrefmark{1}Computing Systems Lab, Huawei Zurich Research Center, Switzerland \\
      \IEEEauthorrefmark{2}Barcelona Supercomputing Center, Spain}
      \thanks{Address correspondence to renzo.andri@huawei.com}
      }

\maketitle

\begin{abstract}\label{sec:abstract}
Modern DNN workloads increasingly rely on activation functions consisting of computationally complex operations. This poses a challenge to current accelerators optimized for convolutions and matrix-matrix multiplications. This work presents \prj{}, a lightweight hardware accelerator for activation functions implementing non-uniform piecewise interpolation supporting multiple data formats. Non-Uniform segments and floating-point numbers are enabled by implementing a binary-tree comparison within the address decoding unit. An SGD-based optimization algorithm with heuristics is proposed to find the interpolation function reducing the mean squared error. Thanks to non-uniform interpolation and floating-point support, \prj{} achieves on average \sanity{22.3x} better mean squared error compared to previous piecewise linear interpolation approaches. The evaluation with more than 700 computer vision and natural language processing models shows that \prj{} can, on average,  improve the end-to-end performance of state-of-the-art AI hardware accelerators by 35.7\%, achieving up to 3.3x speedup with negligible impact in the models' accuracy when using 32 segments, and only introducing an area and power overhead of 5.9\% and 0.8\% relative to the baseline vector processing unit.

\end{abstract}
\glsresetall

\setcounter{footnote}{1}

\footnotetext{Both authors contributed equally to this research. This work was done during Enrico Reggiani's internship at Huawei Zurich Research Center.}


\section{Introduction}\label{sec:introduction}

\begin{figure}[t!]
  \centering   
  \includegraphics[width=0.49\textwidth]{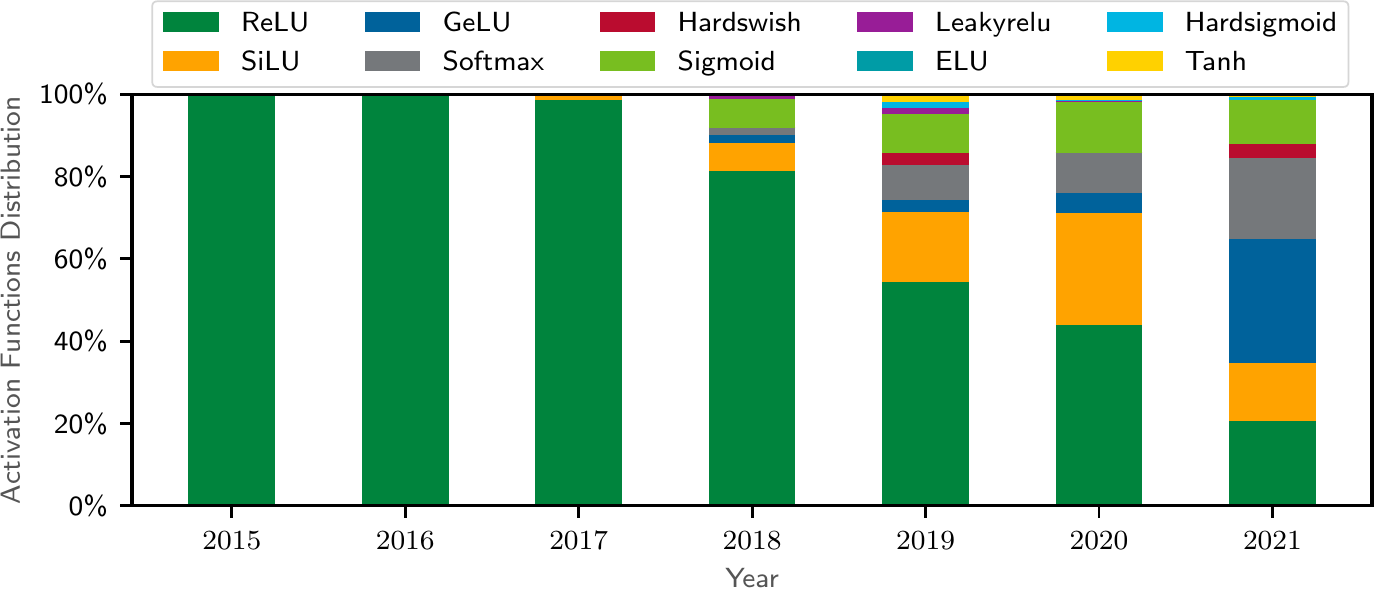}
  \caption[]{Activation functions distribution by year of model publication, extracted from 700+ \gls{soa} \gls{ai} models of the TIMM and Hugging Face collections \cite{timm,huggingface}. 
  }
  \label{fig:acts_distr}
\end{figure}

To keep pace with the evolution of \gls{ai} models, industry and academia are exploring novel hardware architectures, featuring heterogeneous processing units capable of achieving orders of magnitude improvements in terms of performance and energy efficiency with respect to general-purpose processors.
As the execution time of state-of-the-art \glspl{dnn} has been dominated by operations like convolutions and matrix-multiplications \cite{resnet, vgg}, \gls{dnn} hardware accelerators currently allocate most of their computational resources to specialized linear algebra cores, while leaving the execution of the other layers to general-purpose \glspl{vpu} \cite{tpu_paper, davinci}.
However, aiming at reducing training and inference time and enabling deployment on IoT and edge devices, recent deep learning research efforts are pushing towards decreasing the models' dimension while providing comparable or better accuracy.
To this aim, recent networks increase their heterogeneity by introducing new layers featuring reduced operational intensity and new parameter-free layers.
As these new layers can be efficiently computed by exploiting highly data-parallel accelerators, they are significantly increasing the share of execution time spent on the \glspl{vpu} hosted in \gls{dnn} hardware accelerators.
Specifically, parameter-free layers like the \relu{} are increasingly replaced with activation functions requiring a higher compute effort, such as the \gelu{}, the \silu{}, and \texttt{Softmax}, which are composed of several expensive operations like divisions and exponentiations \cite{act_survey}.
This trend can be seen clearly in \Cref{fig:acts_distr}, analyzing the activation functions distribution over the past years, considering $628$ computer vision \glspl{dnn} and $150$ \gls{nlp} transformers from \textit{PyTorch Image Models (TIMM)} \cite{timm} and \textit{Hugging Face} \cite{huggingface}, respectively. 
While \relu{} was the dominant activation function from 2015 to 2017, it declined to $20.7\%$ in 2021, while functions like \silu{} and \gelu{} emerged over the last years, jointly accounting for $32.1\%$ and $44.2\%$ of the total activation functions count in 2020 and 2021, and requiring $4\times$ and $12\times$ more arithmetic operations than \relu{}, respectively.
Therefore, exploring novel hardware architectures capable of efficiently accelerating the computation of complex activation functions is becoming an increasingly important research topic.

In this paper, we propose \prj{}, a flexible hardware accelerator for deep learning activation functions, integrated in general-purpose \glspl{vpu} and relying on a novel \gls{pwl} approximation approach, capable of averagely improving the precision of previous \gls{pwl} approximation approaches by \sanity{$22.3\times$}.
The main novelty of \prj{} relies on its flexibility.
Indeed, its functionality can be reprogrammed to approximate all common activation functions. It supports 8-, 16-, and 32-bit fixed-point and floating-point data formats, and it allows selecting arbitrary locations for the \gls{pwl} interpolation points. 

The main contributions of this paper are listed as follows:
\begin{itemize}
    \item We propose a reprogrammable hardware architecture, extending the set of functional units hosted in \glspl{vpu}, capable of improving the performance of activation functions computations and supporting both fixed-point and floating-point operations;
    \item We define a \gls{pwl} algorithm to automatically find the best interpolation points of any \gls{dnn} activation function, supporting arbitrary positions of the interpolation points and achieving negligible top-1 accuracy drops on the analyzed models considering 16 or 32 \textit{breakpoints};
    \item An end-to-end performance and accuracy evaluation of \prj{}, targeting a commercial \gls{dnn} accelerator and benchmarking more than 700 \gls{soa} deep learning models.
\end{itemize}


\section{Background and Motivation}\label{sec:related}

Activation functions represent one of the most common and important layers of \glspl{dnn}, as they apply non-linear transformations to the network feature maps.
While \relu{} has been widely used for many deep learning tasks, modern networks are using more complex activation functions \cite{af_survey1} to achieve higher accuracies and avoid the well-known \textit{``dying \relu{} effect''} \cite{dying_relu}.
As these kernels require many complex operations, they are typically accelerated via \textit{function approximation} strategies, whose methods can be grouped into three main categories: \textit{polynomial}, \textit{\gls{lut}-based}, and \textit{hybrid}.

\textit{Polynomial} approximation methods \cite{taylor, efficient_vlsi} compute the activation functions through series expansions, such as Taylor and Chebyshev approximations. 
Although these methods feature high-precision computations, their hardware implementations are typically tailored to a specific activation function and are costly in terms of area, as their computation requires several \gls{madd} operations.

\textit{\gls{lut}-based} architectures \cite{andri2020extending, an_optimized, high_speed_vlsi, twofold} feature higher flexibility than \textit{polynomial} methods.
They perform a \gls{pwl} approximation, subdividing the function input range into \textit{intervals} and associating each \textit{interval} to a specific function output, whose value is pre-computed and stored in memories used as \glspl{lut}.
An addressing scheme is then used to map a given input \cite{an_optimized} or a full \textit{interval} \cite{high_speed_vlsi} to a specific \gls{lut} address, holding the corresponding function output. 
\textit{\gls{lut}-based} solutions can be more flexible than \textit{polynomial} methods, as programmable \glspl{lut} can store different sets of output data depending on the target activation function. 
However, they require a high area footprint to provide good accuracy.
Indeed, as the function output is directly provided by the \gls{lut}, the approximation precision strictly depends on the number of \textit{intervals} (\ie{} the \gls{lut} depth) in a selected input range.

To overcome these limitations, several works \cite{an_efficient_2006, an_efficient_2018, low_cost, hw_based, low_overhead} explore \textit{hybrid} solutions, which combine the \textit{polynomial} and \textit{\gls{lut}-based} approaches by storing in the \glspl{lut} the interpolation \textit{segment} coefficients instead of the function outputs.  
For example, a \gls{pwl} \textit{hybrid} approach approximates a given activation function as $N$ straight lines (\ie{} \textit{segments}), each satisfying the equation $f(x) = m_ix+q_i$, for $i \in \{0, 1, ..., N-1\}$.
A \gls{madd} operation is then used to compute the function output (\ie{} \textit{f(x)}) starting from a specific set of \textit{segment} \textit{coefficients} stored in the \glspl{lut} (\ie{} \textit{$m_i$, $q_i$}) and from the incoming input data (\ie{} \textit{x}). 
\begin{figure}[t!] 
    \centering   
    \includegraphics[width=0.47\textwidth]{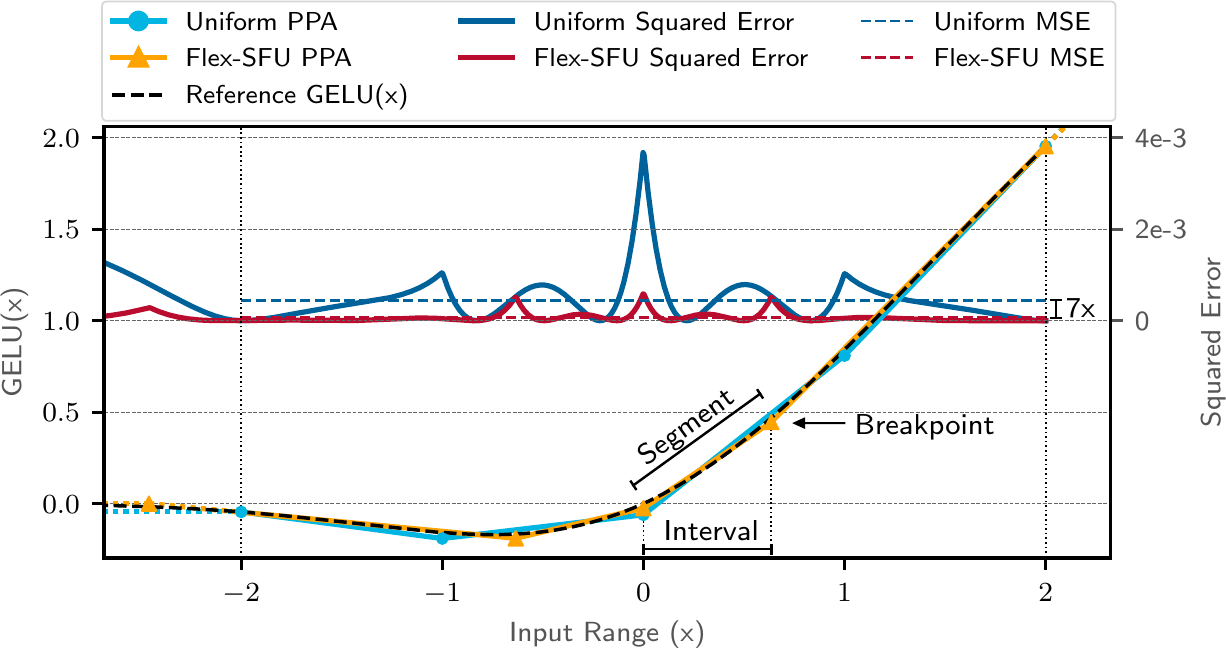}
    \caption[]{\gls{pwl} approximation (left \textit{y}-axis) and squared error (right \textit{y}-axis) of \gelu{}, exploiting uniform and non-uniform (\ie{} \prj{}) interpolations, and considering 5 \textit{breakpoints} (\ie{} 4 \textit{segments}) in the [-2,2] input range.}
    \label{fig:gelu_approx} 
\end{figure}
\begin{figure*}[t!] 
    \centering    
    \includegraphics[width=1\textwidth]{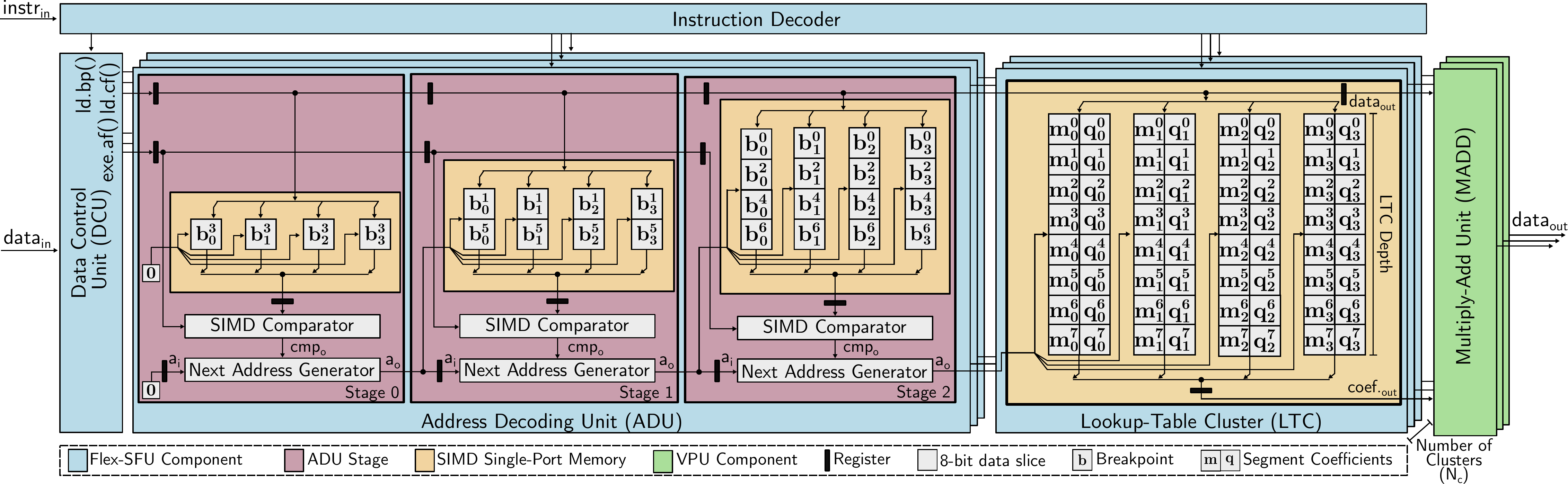}
    \caption[]{\prj{} hardware architecture, integrated into the main \gls{vpu} as an additional functional unit, considering Look-up table clusters (LTCs) capable of storing 8 \textit{segment} coefficients. Memory cell superscripts represent \textit{breakpoints} and \textit{coefficients} IDs, while memory cell subscripts represent data slices.}
    \label{fig:architecture}
\end{figure*}
\textit{Hybrid} solutions outperform \textit{\gls{lut}-based} approaches in terms of area and accuracy, as they are able to correctly approximate the whole \textit{segment} instead of selecting a reference output value for a given \textit{interval}.
Moreover, \textit{hybrid} approaches relax the constraint on the maximum function input range, as they allow to approximate any function featuring boundaries that converge to a fixed slope.

However, current \textit{hybrid} solutions present several limitations.
\ding{182} They are tailored to a single input data type, either converted into a fixed-width fixed-point notation \cite{an_efficient_2006, an_efficient_2018, low_cost, plac} or only considering a single floating-point format \cite{hw_based, low_overhead}.
Moreover, their \gls{lut} addressing schemes simply rely on a fixed subset of bits, such as the input data \glspl{msb}. 
However, these approaches lack flexibility, firstly because current accelerators support several data formats and \gls{simd} computations \cite{tpu_comp}, (\eg{} from four 8-bit to one 32-bit elements/cycle), and secondly because their addressing schemes need to be tailored for each target function and input data type. 
\ding{183} Their approximation methods mainly rely on uniform interpolations (\ie{} \textit{segments} share the same length).
However, activation function approximations would benefit from non-uniform interpolation granularity among different function \textit{intervals}, as it would allow increasing the density of \textit{segments} on more sensitive \textit{intervals} while relaxing their density on straight \textit{intervals}.
For example, in \Cref{fig:gelu_approx}, showing the \gls{pwl} approximation of \gelu{} exploiting both uniform and non-uniform interpolations, the non-uniform strategy improves the \gls{mse} by $7 \times$ while using the same number of \textit{breakpoints}.
Although non-uniform strategies exist in the literature, they either rely on simply removing \textit{breakpoints} from a uniform interpolation while maintaining similar precisions \cite{nonuniform0, nonuniform1}, or only optimize the interpolation error for narrow input ranges, leaving it diverging outside the selected interval with unknown impact on the end-to-end accuracy \cite{plac}.
\ding{184} Although many related works\cite{an_efficient_2018, low_cost, hw_based, low_overhead} perform end-to-end accuracy evaluations on selected deep learning models, none of them quantify the accuracy impact of their solutions on a large set of networks.
However, such analyses are crucial to verify the robustness of a given approximation method, as different models can suffer from varying sensitivities to activation function errors.

To overcome these limitations, we propose \prj{}, whose hardware architecture (detailed in \Cref{sec:hw_arch}) supports all the data sizes typically used by \glspl{dnn}, and features linear throughput scaling with constant on-chip memory usage.
Our reprogrammable hardware architecture implements an addressing scheme supporting non-uniform \textit{segments}, whose optimal lengths minimizing the approximation \gls{mse} are determined by a novel \gls{pwl} algorithm, described in \Cref{sec:algo_eval}. 
 
\section{\prj{} Hardware Architecture}\label{sec:hw_arch}

The proposed hardware architecture implements a \textit{hybrid} \gls{pwl} approach to approximate activation functions. It supports both fixed-point and floating-point data formats composed of 8-, 16-, and 32-bit, representing the most used data types targeting deep learning applications.

Depending on the input data value, \prj{} provides the proper \textit{coefficients} to the \gls{vpu} \gls{madd} units, computing the activation function output.
Differently from the related work analyzed in \Cref{sec:related}, performing the address decoding exploiting a subset of incoming input data bits, \prj{} relies on small memories to store the \textit{breakpoint} values on-chip, and compares them with the incoming data to find the respective \gls{lut} address.
This feature allows for higher flexibility than solutions only supporting uniform \textit{segments}, as it permits to select the best length for each \textit{segment}, thus minimizing the approximation error.
  
\prj{} extends the set of functional units available on current \glspl{vpu} targeting deep learning, acting as a \gls{sfu} capable of accelerating activation functions via \gls{pwl} approximation. 
Its execution is handled by three custom instructions extending the target \gls{vpu} \gls{isa}, namely \texttt{ld.bp()}, \texttt{ld.cf()}, and \texttt{exe.af()}.
The proposed instructions are decoded by the \gls{vpu}, and then handled by \prj{}, whose main architectural components are depicted in \Cref{fig:architecture}.
A \gls{dcu} dispatches input data among the other \prj{} units.
Specifically, \texttt{ld.bp()} and \texttt{ld.cf()} source data, holding either \textit{breakpoints} or \gls{pwl} \textit{coefficients}, are sent by the \gls{dcu} to the \gls{adu} or the \gls{ltc} unit, and stored in \textit{\gls{simd} single-port memories}.
These instructions must be executed only once when a different activation function has to be computed, and can be pre-executed while other accelerator compute units (\eg{} the main tensor-unit) are still computing the activation function inputs. 
Therefore, as discussed in \Cref{sec:ppa_eval}, they do not introduce a large overhead in the overall computation.
Once \textit{breakpoints} and \gls{lut} \textit{coefficients} have been loaded in the \gls{adu} and \gls{ltc} units, multiple \texttt{exe.af()} can be executed to compute the activation function outputs. 
These operations are handled by the \gls{dcu}, which streams the input data through the pipeline composed of the \gls{adu} and the \gls{ltc}.
As \Cref{fig:architecture} shows, the \gls{adu} functionality resembles a \gls{bst}. 
Each \gls{adu} stage defines a \gls{bst} level, and exploits \textit{\gls{simd} single-port memories} to implement \gls{bst} nodes holding \textit{breakpoints}, which are ordered to allow traversing one \gls{bst} level per stage to search for the proper \gls{ltc} address depending on the input data.
Each cycle, a \textit{\gls{simd} comparator} supporting both fixed-point and floating-point number formats determines if the current input data is greater or smaller than the \textit{breakpoint} loaded from memory exploiting the \textit{cmp\textsubscript{o}} signal, whose value can be either 1 or 0, respectively.
The comparison output and the input address are then used by the \textit{Next Address Generator} unit to find the subsequent \gls{adu} stage address, namely $a_o$.  
The last \gls{adu} stage performs the comparison among the \gls{bst} leaves, thus finding the proper \gls{lut} address which is forwarded to the \gls{ltc} unit.
Finally, the \gls{ltc} loads the appropriate \textit{segment} \textit{coefficients}, and sends them and the delayed input data to the \gls{vpu} \gls{madd} functional units, computing the activation function output.

The memory-mapping strategy exploited by the \gls{adu} and \gls{ltc} units consists of four \textit{\gls{simd} single-port memories}, whose bit-width is equal to the product between the minimum supported bit-width (\ie{} 8-bit) and the number of \textit{coefficients} (\ie{} 1 and 2 for the \gls{adu} and \gls{ltc}, respectively).
Each memory is accessed separately in case of computations based on 8-bit data (\eg{} $b^i_0$, $b^i_1$, $b^i_2$, $b^i_3$ are accessed as four separate 8-bit data), while for 16-bit computations each data is segmented among two subsequent memories (\eg{} $b^i_0$ - $b^i_1$ and $b^i_2$ - $b^i_3$ are accessed as two 16-bit data), in such a way to support an input throughput of two 16-bit elements/cycle.
Similarly, the same data is partitioned among the four 8-bit memories in case of 32-bit computations (\eg{} $b^i_0$ - $b^i_1$ - $b^i_2$ - $b^i_3$ are accessed as a single 32-bit data), allowing to support a throughput of one 32-bit element/cycle, while reusing the same memories.

As shown in \Cref{fig:architecture}, to allow for further scalability, the \prj{} parallelism can be tuned by increasing the number of instantiated clusters, namely \textit{N\textsubscript{c}}, to match the underlying \gls{vpu} throughput.
Note that, as \glspl{vpu} are typically optimized for throughput, we design \prj{} exploiting pipelining, thus enabling steady-state performance of \textit{N\textsubscript{c}} $\times$ 32-bit/cycle, while avoiding dead-locks by design.


\section{\prj{} Approximation Methodology}\label{sec:algo_eval}

\begin{figure}[t!]
    \centering   
    \includegraphics[width=0.46\textwidth]{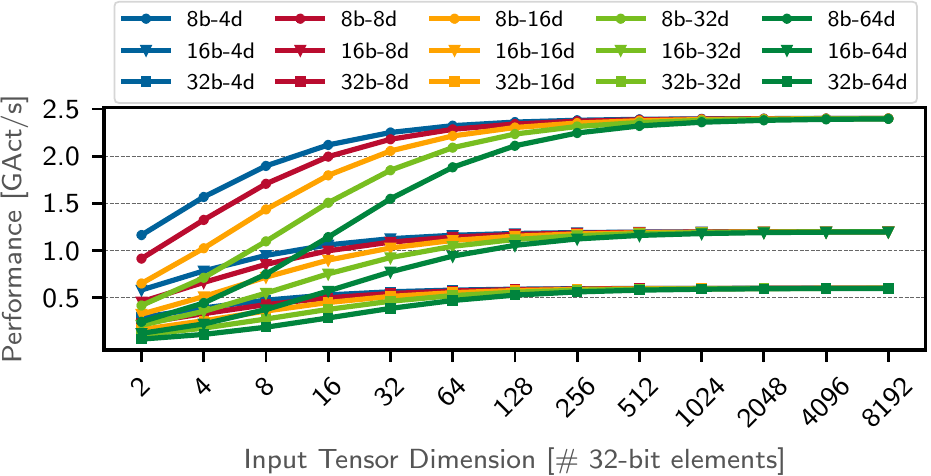}
    \caption[]{Throughput of \prj{}, in terms of number of computed activations (\ie{} \textit{GAct}) per second, as a function of the input tensor size, accounting for different bit-widths (\textit{b}) and lookup table cluster (LTC) depths (\textit{d})}
    \label{fig:sfu_performance}
\end{figure}

We rely on a \gls{pwl} approximation, defining the interpolated and steady function $\hat f (x)$ as:
 
\[f(x) \approx \hat f (x) =
\begin{cases}
m_{l}(x-p_0)+v_0 & x \leq p_0 \\
\frac{v_{i+1}-v_{i}}{p_{i+1}-p_{i}}(x-p_i)+v_i & p_i < x < p_{i+1},\\ & 0<i<n-1 \\
m_{r}(x-p_{n-1})+v_{n-1} &  x \geq p_{n-1}
\end{cases}
 \]
 
\vspace{10pt}
 with $n$ \textit{breakpoints} $p_i$, $(n+1)$ linear \textit{segments}, and $n$ function values at the \textit{breakpoints} $v_i=\hat f(p_i)$. 
 The most left and right \textit{segments} are calculated with values $v_0$ and $v_{n+1}$, using slopes $m_l$ and $m_r$, while the inner \textit{segments} of each \textit{breakpoint} $p_i$ are linearly interpolated through its value $v_i$ and the following \textit{breakpoint}-value pair [$p_{i+1}$;$v_{i+1}$]. 

 To find the \textit{breakpoint}-value pairs, we start with uniformly distributed \textit{breakpoints} and exact function values. We use the Adam optimizer \cite{adam} (with lr=0.1, momenta=(0.9, 0.999)) and the Plateau LR scheduler. We choose the \gls{mse} between the interpolated function $\hat f$ and the target function $f$ on the interval $[a,b]$ as the loss function:
 \[
 \mathcal{L}_{[a,b]}(\hat f, f) = \frac{1}{b-a}\int_a^b (\hat f(x) - f(x))^2 dx
 \]

Aiming to avoid stalls in sub-optimal local minima during the optimization process, we extend our optimization algorithm by removing \textit{breakpoints} and reinserting them at a better location.

\textit{Removal loss:} 
We define the removal loss $\ell_i^\text{rm}$ as the loss of the interpolated function if the \textit{breakpoint} $p_i$ is removed. We then remove the \textit{breakpoint} with the minimal removal loss, $p_\mathrm{remove}$:

\begin{align*}
p_\mathrm{remove}=\argmin_{p_i}\;\ell_i^\text{rm}, \qquad\ell_i^\text{rm}=\mathcal{L}_{[a,b]}(\hat f_{{\{p_j, v_j\}}_{j \ne i}}, f).
\end{align*}

\textit{Insertion loss:} 
On the other hand, we define the insertion loss $\ell_i^\text{ins}$ as the loss over the $i$-th \textit{segment}, and insert a new \textit{breakpoint} in the center of the \textit{segment} with the highest insertion loss:

\begin{align*}
\begin{pmatrix}p_\mathrm{insert}\\v_\mathrm{insert}\end{pmatrix}=\argmax_{\scriptsize\begin{pmatrix}(p_i+p_{i+1})/2\\(v_i+v_{i+1})/2\end{pmatrix}} \ell_i^\text{ins},\  \ell_i^\text{ins}=(p_{i+1}-p_i)\mathcal{L}_{[p_i,p_{i+1}]}(\hat f, f).
\end{align*}

\textit{Boundary condition}: All relevant activation functions converge outside the interpolation interval to a constant value or an asymptote. To avoid large errors outside of the interpolation interval, unless noted otherwise, we define boundary conditions for value and slope for the most left and the most right \textit{segments}, such that they lie on the asymptote of the function:
\begin{align*}
m_l&=\lim_{x\to -\infty}f(x)/x, &v_0 &= m_lp_{0}+\lim_{x\to -\infty}(f(x)-m_l x) \mathrm{,} \\
m_r&=\lim_{x\to +\infty}f(x)/x, &v_{n-1} &= m_rp_{n-1}+\lim_{x\to +\infty}(f(x)-m_r x)
\end{align*}
For example, considering \gelu{}, this resolves to $m_l=0, v_0=0, m_r=1, v_{n-1}=p_{n-1}$. Notably, $p_0$ and $p_{n-1}$ themselves are still learned. In this way, the interpolated function converges to the original function for values far from the boundary \textit{breakpoints}. 
This comes at a small cost in error close to the boundary \textit{breakpoints}.

\textit{Optimization strategy:} We initialize the \prj{} function interpolation with uniformly distributed \textit{breakpoints}. Then we optimize with SGD until convergence. After this, we remove and insert one \textit{breakpoint} as described above, and retrain with a lower learning rate. We reiterate until removal and insertion points converge. Note that we perform this optimization for each function, and we substitute the layers within the \gls{dnn} models without any retraining for ease of use.

\begin{table}[t!]
    \centering
    \caption{\prj{} Characterization for \textit{N\textsubscript{c}} = 1 at $f=600$\,MHz in 28nm CMOS}
    \resizebox{0.98\columnwidth}{!}{%
    \scriptsize
    \begin{tabular}{lrrrrr}
    \toprule
    {\textbf{\gls{ltc} Depth (\ie{} \# Segments)}}    & \textbf{4}  & \textbf{8} & \textbf{16} & \textbf{32} & \textbf{64}  \\
    \midrule
    \textbf{Latency [cycles]} & 7 & 8 & 9 & 10 & 11 \\
    \textbf{Power [mW]}    & 1.4 & 1.7 & 2.2 & 2.8 & 3.7 \\
    \midrule
    \textbf{\gls{adu} Area {[}\%{]}} & 34.2\% & 41.2\% & 43.7\% & 46.0\% & 41.6\%  \\
    \textbf{\gls{ltc} Area {[}\%{]}} & 31.3\% & 34.9\% & 44.1\% & 46.6\% & 53.4\%  \\
    \textbf{Total Area     {[}\textmu{}m\textsuperscript{2}{]}} & \textbf{2572.4} & \textbf{3593.0} & \textbf{5846.0} & \textbf{9791.3} & \textbf{14857.2} \\
    \bottomrule  
    \end{tabular} 
    }{}
    \label{tab:characterization}
\end{table} 

\section{Experimental Evaluation}\label{sec:results}

\subsection{Performance, Power and Area Analyses}\label{sec:ppa_eval}

We implement the proposed hardware accelerator in \gls{rtl}, and perform synthesis and \gls{pnr} for a 28nm CMOS technology node.
We evaluate several \prj{} configurations in terms of performance, area, and power, varying the number of \textit{segments} from 4 to 64 while considering $N_c=1$ and a target frequency of 600\,MHz.
\Cref{fig:sfu_performance} shows the throughput of \prj{}, accounting for the time spent on both \texttt{ld.bp()}, \texttt{ld.cf()}, and \texttt{exe.af()}, across input tensors ranging from 2 to 8k 32-bit data.
All the analyzed \prj{} combinations reach the steady-state performance for input tensors larger than 256 32-bit data, gaining 0.6\,GAct/s, 1.2\,GAct/s, and 2.4\,GAct/s in terms of throughput, considering 8-, 16-, and 32-bit data sizes, corresponding to an energy efficiency ranging from 158\,GAct/s/W to 1722\,GAct/s/W.
Note that the throughput reported in \Cref{fig:sfu_performance} saturates to 1\,OP/cycle, 2\,OP/cycle, and 4\,OP/cycle for 32-, 16-, and 8-bit data sizes at 600 MHz, proving that \prj{} can reach the theoretical peak performance discussed in \Cref{sec:hw_arch}, accelerating complex \gls{dnn} activation functions exploiting the same computation time typically required by simple operations like \relu{}.

\Cref{tab:characterization} details the characterization of \prj{}, obtained after the \gls{pnr} step and considering from 4 to 64 \textit{segments}, reporting a total power consumption ranging from 1.4\,mW to 3.7\,mW, and a total area requiring from 2572\,\textmu{}m\textsuperscript{2} to 14857\,\textmu{}m\textsuperscript{2}.

To investigate the area and power impact of \prj{} on high-performance \glspl{vpu}, we perform a back-of-the-envelope integration of \prj{} into the RISC-V \gls{vpu} proposed by Perotti et al. in \cite{ara}, composed of 4 lanes and supporting a maximum data size of 64-bit.
Our evaluation, considering four \prj{} instances (\ie{} one instance per lane) featuring $N_c=2$ (\ie{} supporting from $1\times$64-bit to $8\times$8-bit elements/cycle), shows that \prj{} only accounts for 2.2\%, 3.5\% and 5.9\% of the total area for a \gls{ltc} depth of 8, 16 and 32 elements, respectively, while consuming from 0.5\% to 0.8\% of the total power.

\subsection{Function Approximation Precision Analysis}\label{sec:prec_eval}

\begin{figure}[t!]  
    \centering     
    \includegraphics[width=0.48\textwidth]{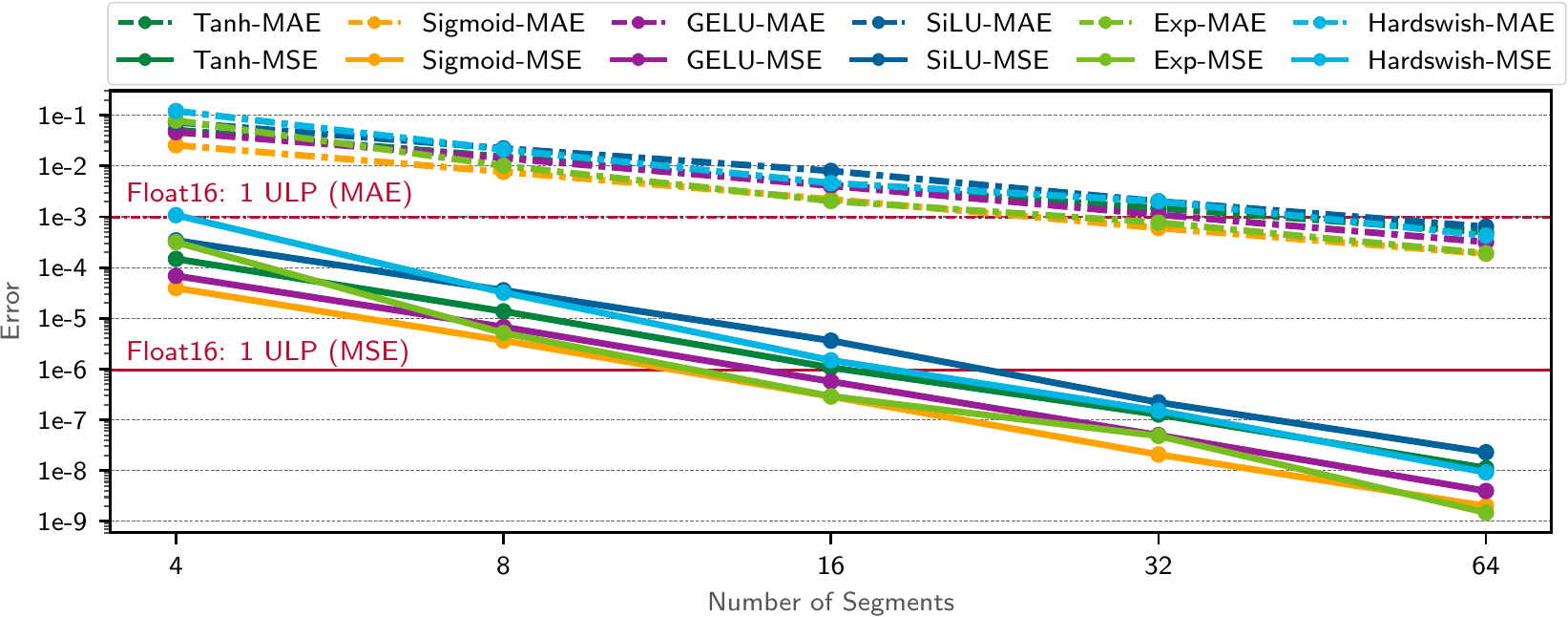}
    \caption[]{Error analysis for a set of activation functions, considering from 4 to 64 \textit{breakpoints}. Interpolation intervals are in range [-10, 0.1] for the exponential function (\texttt{Exp}), and in range [-8, 8] for the other functions.}
    \label{fig:sfu_precision}
\end{figure}

In \Cref{fig:sfu_precision}, we investigate \gls{mse} and \gls{mae} of the most representative activation functions. We select the interpolation interval within [-10,0.1] for \texttt{Exp}, and within [-8,8] for the other functions. 
The boundary \textit{breakpoints} lie on the functions' asymptote to reduce the error outside the interpolation interval. 
We interpolate \texttt{Exp} for negative values to be used in \texttt{Softmax}, typically requiring exponentiation implemented with (vector-wide) maximum subtraction (\ie{} $\exp(x_i-\max_j\;x_j)$).
As detailed in \Cref{fig:sfu_precision}, the approximation precision of the analyzed functions averagely improves \gls{mse} and \gls{mae} by $15.9\times$ and $3.8\times$ per doubling of the number of \textit{breakpoints}.
Moreover, all the interpolations featuring more than 16 \textit{breakpoints} reach a \gls{mse} lower than 1 Float16 \gls{ulp}, defined as the single-bit error at a base of 1.

In \Cref{tab:soaerror}, we compare \prj{} with other \gls{pwl} interpolation methods, considering the same interpolation range and number of \textit{breakpoints}. 
Following most of the previous works, \cite{an_efficient_2006,an_efficient_2018,low_cost,hw_based,low_overhead}, we evaluate \prj{} exploiting the \gls{aae} metric, squaring it (\ie{} sq-AAE) to match the same \gls{mse} order of magnitude. 
Furthermore, we compare with the equivalent number of \textit{breakpoints} of previous works exploiting symmetry \cite{an_efficient_2006,andri2020extending}.
As \Cref{tab:soaerror} shows, our method outperforms all the other \gls{pwl} approaches, by a factor ranging from 2.3$\times$ to \sanity{88.4$\times$}, with an average of 22.3$\times$.

The method proposed by Gonzalez et al. \cite{hw_based} exploits a second-order piecewise but not-steady interpolation, averagely achieving $4.3\times$ better \glspl{mse} than \prj{} on \texttt{Tanh}, \texttt{Sigmoid}, and \silu{}.
Although \prj{} can be easily extended to support a second-order interpolation, we believe that the \gls{pwl} methodology of \prj{} represents a better candidate to compute activation functions on \gls{dnn} accelerators.
Indeed, second-order approximations feature high area overheads, requiring to double the number of \gls{vpu} \gls{madd} units to guarantee the same throughput of the proposed solution, as well as larger \glspl{lut} able to store an additional interpolation coefficient. 

\begin{table}[t!]
    \centering
\caption{Comparison of our \gls{mse}-optimized method with other PWL Interpolation Methods with the same number of \textit{breakpoints} and range}
\label{tab:soaerror}
\begin{threeparttable}
    \resizebox{0.88\columnwidth}{!}{    
\begin{tabular}{cclcllr}
\toprule
& \multicolumn{3}{c}{\textbf{Parameters}} & \multicolumn{3}{c}{\textbf{Error sq-AAE}\tnote{*}} \\
\cmidrule(lr){2-4}\cmidrule(lr){5-7}
        & \textbf{Funct.} & \multicolumn{1}{c}{\textbf{Range}} & \textbf{\#BP} &\multicolumn{1}{c}{\textbf{Ref.}} &\multicolumn{1}{c}{\textbf{This work}} &   \multicolumn{1}{c}{\textbf{Impr.}} \\ 
        \midrule
 \cite{an_efficient_2006}  & \multirow{6}*{\textbf{Tanh}} & [-8, 8]     & 16\tnote{\textdagger} & $5.76\cdot10^{-6}$	&	$4.27\cdot 10^{-7}$	&	13.5$\times$\\
 \cite{an_efficient_2018}  &  & [-3.5, 3.5] & 16                                                & $3.58\cdot 10^{-5}$	&	$1.52\cdot 10^{-6}$	&	23.5$\times$\\
 \cite{an_efficient_2018}  &  & [-3.5, 3.5] & 64                                                & $1.12\cdot 10^{-7}$	&	$7.88\cdot 10^{-9}$	&	14.2$\times$\\

 \cite{low_cost}           &  & [-8, 8]     & 16                                                & $1.00\cdot 10^{-6}$	&	$4.26\cdot 10^{-7}$	&	2.3$\times$\\
 \cite{low_overhead}       &  & [1/64, 4]   & 32                                                 &$5.94\cdot 10^{-7}$	&	$6.72\cdot 10^{-9}$	&	88.4$\times$\\

 \cite{andri2020extending} &  & [-4, 4]     & 32\tnote{\textdagger}                              &$9.81\cdot 10^{-7}$\tnote{\ddag}	&	$1.13\cdot 10^{-8}$\tnote{\ddag}	&	86.8$\times$\\

\midrule
 \cite{an_efficient_2006}  & \multirow{6}*{\textbf{Sigmoid}} & [-8, 8]     & 16\tnote{\textdagger} & $8.10\cdot 10^{-7}$	&	$1.21\cdot 10^{-7}$	&	6.7$\times$\\

 \cite{an_efficient_2018}  &  & [-7, 7]     & 16                                                    & $8.95\cdot 10^{-6}$	&	$4.97\cdot 10^{-7}$	&	18.0$\times$\\
 \cite{an_efficient_2018}  &  & [-7, 7]     & 64                                                    & $2.82\cdot 10^{-8}$	&	$2.38\cdot 10^{-9}$	&	11.9$\times$\\
 \cite{low_cost}           &  & [-8, 8]     & 16                                                    & $6.25\cdot 10^{-6}$	&	$2.88\cdot 10^{-7}$	&	21.7$\times$\\

 \cite{low_overhead}       &  & [1/64, 4]   & 32                                                    & $1.41\cdot 10^{-7}$	&	$3.80\cdot 10^{-8}$	&	3.7$\times$\\

 \cite{andri2020extending} &  & [-4, 4]     & 64\tnote{\textdagger} & $3.92\cdot 10^{-8}$\tnote{\ddag} & $2.38\cdot 10^{-9}$\tnote{\ddag} & 9.3$\times$ \\
\midrule
\cite{low_cost}           & \textbf{GeLU}    & [-8, 8]     & 16  & $6.76\cdot 10^{-6}$ & $1.89\cdot 10^{-7}$ & 9.0$\times$\\
\bottomrule
\end{tabular}
}{}
\begin{tablenotes}[para]
    \footnotesize
      \item[*] SoA reports average absolute error (AAE).
      \item[\ddag] Numbers in \gls{mse}. 
      \item[\textdagger] Uses symmetry to halve the number of \textit{segments}. 
    \end{tablenotes}
  \end{threeparttable}
  \vspace{-6mm}
\end{table}
\subsection{End-to-End Evaluation}\label{sec:e2e_eval} 

We evaluate \prj{} on a commercial \textit{Huawei Ascend 310P} \gls{ai} processor \cite{ascend}, exploiting a benchmark suite targeting $628$ computer vision and $150$ \gls{nlp} networks from \textit{PyTorch Image Models (TIMM)} and \textit{Hugging Face}, respectively.
This accelerator represents an ideal candidate to demonstrate the benefits that \prj{} can provide to \gls{soa} \glspl{dnn} accelerators, as it hosts a specialized matrix multiplication unit computing up to 4096\,MAC/cycle, and processes the \gls{dnn} activation functions on a general-purpose high-performance \gls{vpu}.
To perform our performance evaluation, we convert each benchmark suite model from Pytorch~1.11 \cite{paszke2019pytorch} to ONNX~1.12 \cite{onnx} with \textit{opset} version 13, and we replace each activation function of the resulting model graph with a custom ONNX operator, implementing a set of instructions supported by the \textit{Huawei Ascend} \gls{isa}, and whose latency and throughput match the \prj{} metrics presented in \Cref{sec:ppa_eval}.
Then, we compile both the baseline and the \prj{}-based ONNX models for the \textit{Ascend} \gls{ai} processor with \gls{atc}~v5.1 and run them on the target accelerator to extract and compare their end-to-end inference run time.
In our evaluation, we compute each model using all 8 cores of the \textit{Ascend 310P} \gls{ai} processor in parallel with batch size equal to 1, considering the average execution time between 10 subsequent inference runs.

\begin{figure}[t!]
    \centering
    \includegraphics[width=0.5\textwidth]{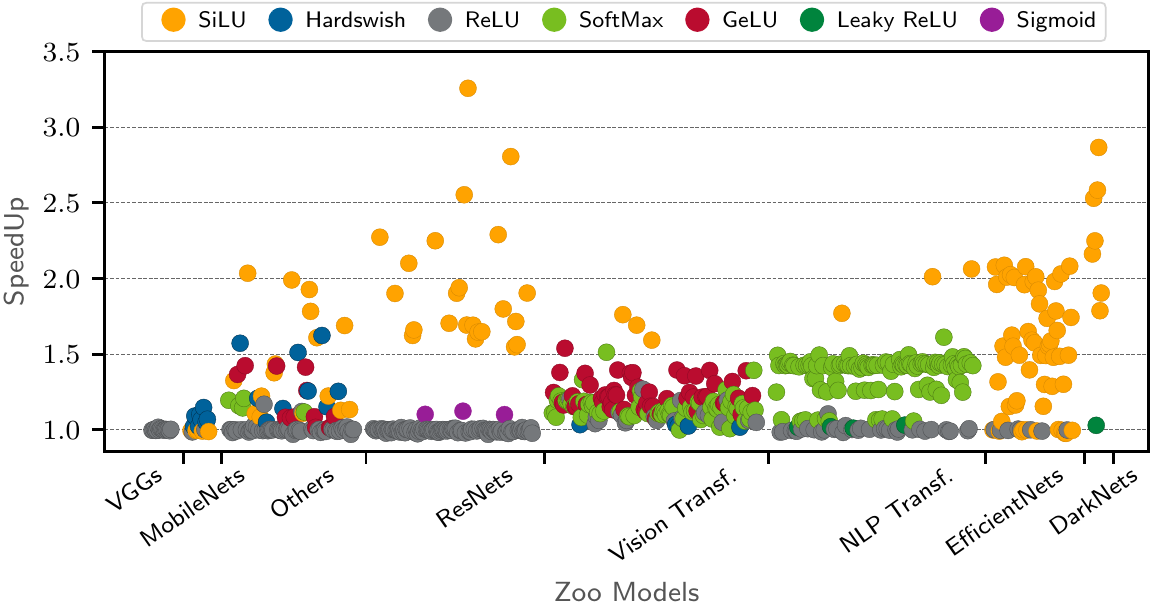}
    \caption{End-to-end model zoo performance evaluation on the \textit{Huawei Ascend 310P} \gls{ai} processor. Colors: Most frequently used activation function in DNN.}
    \label{fig:zoo_speedup}
\end{figure}

\Cref{fig:zoo_speedup} summarizes the execution time improvements of the proposed benchmark suite when exploiting \prj{}, highlighting the reference family and most frequent activation function of each model.
We obtained comparable performance results, not reported for space reasons, for batch sizes equal to 16, 32, and 128.
As \Cref{fig:zoo_speedup} shows, \prj{} matches the performance of models primarily relying on lightweight activation functions (\ie{} \relu{}, \texttt{Leaky ReLU}), not introducing any overhead in their computation, and greatly improves the execution time of networks relying on more complex activation functions.
Specifically, including the models based on \relu{}, whose baseline execution time matches the \prj{} performance, \prj{} allows gaining $17.3\%$, $17.9\%$, $29.0\%$, and $45.1\%$ performance on \textit{ResNets}, \textit{Vision Transformers}, \textit{\gls{nlp} Transformers}, and \textit{EfficientNets} models, while reaching $2.1\times$ more performance on \textit{DarkNets} models.
Overall, \prj{} reaches $22.8\%$ better performance on the considered model zoo computation, improving the execution time of models relying on complex activation functions by $35.7\%$ on average, and reaching a performance peak of $3.3\times$ on the computation of \textit{resnext26ts}. 

We evaluate the accuracy impact of \prj{} for \glspl{dnn} in the TIMM database on the ImageNet dataset~\cite{imagenet} by comparing the top-1 accuracy on the validation set between the reference model and one where the activations are replaced with \prj{}.
\Cref{tab:endtoendaccuracy} shows the percentage of networks featuring accuracy drops ranging from 0.1\,\% to 2\,\% with respect to the baseline accuracy, as well as the mean and maximum accuracy drops for 4 to 64 \textit{breakpoints}.
As \Cref{tab:endtoendaccuracy} shows, while only a few networks can tolerate 4 \textit{breakpoints}, 8 and 16 \textit{breakpoints} already allow \sanity{80\%} and \sanity{90\%} of the networks to achieve a drop smaller than 0.1\%, with a mean drop of \sanity{0.87\%} and \sanity{0.26\%}, respectively.
Moreover, 32 and 64 \textit{breakpoints} are almost lossless, with \sanity{99\%} and \sanity{100\%} of the models showing less than 0.1\% drop, and featuring maximum accuracy penalties of \sanity{0.30\%} and \sanity{0.04\%}, respectively.
We noted that networks using \silu{} are the most sensitive to approximation. 
For example, to feature an accuracy drop smaller than 0.17\%, \emph{mobilevit} and \emph{halonet50ts} require 32 \textit{breakpoints}, while \emph{lambda\_resnet50ts} and \emph{mixer\_b16\_224\_miil} require 16 \textit{breakpoints}. 
\texttt{Hardswish} is the second most sensitive activation function, with \emph{lcnet} and \emph{mobilenetv3\_small} requiring 32 \textit{breakpoints}, and \emph{hardcorenas}, \emph{fbnet}, and \emph{mobilenetv3\_large} requiring 16 \textit{breakpoints} to show losses smaller than 0.15\%.  
Finally, \gelu{}-based \emph{sebotnet33ts\_256}, mixer and \emph{crossvit} achieve lossless accuracy drops with 16 \textit{breakpoints}.


\section{Conclusion}\label{sec:conclusion}
We proposed \prj{}, a scalable hardware accelerator for \gls{dnn} activation functions on \glspl{vpu} based on a novel interpolation methodology supporting non-uniform \textit{breakpoints} locations, and performing 8-, 16-, and 32-bit computations based on both fixed-point and floating-point data formats.
Our evaluation shows that \prj{} features low area and power overheads, and reaches \sanity{$22.3 \times$} \gls{mse} improvements on average with respect to other \gls{soa} \gls{pwl} approaches.
Moreover, our end-to-end evaluation on more than 700 \gls{soa} \glspl{dnn} shows that commercial \gls{dnn} accelerators can benefit from \prj{}, allowing them to improve their inference performance up to $3.3\times$ while retaining the models' accuracies.

\begin{table}[t!]
    \centering
\caption{End-to-End Accuracy Drop over 600 DNNs of TIMM\cite{timm}}\label{tab:endtoendaccuracy}
\resizebox{0.99\columnwidth}{!}{%
\begin{tabular}{rrrrrrrrr}
\toprule
& \multicolumn{6}{c}{\textbf{Models Distribution}} & \multicolumn{2}{c}{\textbf{Accuracy Drop}} \\
\cmidrule(lr){2-7} \cmidrule(lr){8-9}
\multicolumn{1}{c}{\textbf{\#BP}} & \multicolumn{1}{c}{\textbf{\textDelta\textless{}0.1}} & \multicolumn{1}{c}{\textbf{\textDelta{\textless{}0.2}}} & \multicolumn{1}{c}{\textbf{\textDelta\textless{}0.5}} & \multicolumn{1}{c}{\textbf{\textDelta\textless{}1}} & \multicolumn{1}{c}{\textbf{\textDelta\textless{}2}} & \multicolumn{1}{c}{\textbf{\textDelta\textgreater{}2}} & \multicolumn{1}{c}{\textbf{mean}} & \multicolumn{1}{c}{\textbf{max}} \\ \midrule

4  & \cellcolor[HTML]{FCFCFF}0.51 & \cellcolor[HTML]{F9FBFC}0.52 & \cellcolor[HTML]{F4F9F8}0.54 & \cellcolor[HTML]{EEF6F3}0.56 & \cellcolor[HTML]{E6F4EC}0.58 & \cellcolor[HTML]{F8696B}0.42 & \cellcolor[HTML]{F8696B}-25.95 & \cellcolor[HTML]{F8696B}-87.00 \\
8  & \cellcolor[HTML]{A2D8B1}0.80 & \cellcolor[HTML]{94D2A5}0.84 & \cellcolor[HTML]{84CC98}0.89 & \cellcolor[HTML]{7CC890}0.92 & \cellcolor[HTML]{72C488}0.95 & \cellcolor[HTML]{FCECEF}0.05 & \cellcolor[HTML]{CBD5AD}-0.87  & \cellcolor[HTML]{F8696B}-77.58 \\
16 & \cellcolor[HTML]{82CB96}0.90 & \cellcolor[HTML]{79C78E}0.93 & \cellcolor[HTML]{72C488}0.95 & \cellcolor[HTML]{6CC283}0.97 & \cellcolor[HTML]{6AC181}0.98 & \cellcolor[HTML]{FCF5F8}0.02 & \cellcolor[HTML]{C8DCB2}-0.26  & \cellcolor[HTML]{F8696B}-25.79 \\
32 & \cellcolor[HTML]{66C07E}0.99 & \cellcolor[HTML]{64BF7C}1.00 & \cellcolor[HTML]{63BE7B}1.00 & \cellcolor[HTML]{63BE7B}1.00 & \cellcolor[HTML]{63BE7B}1.00 & \cellcolor[HTML]{FCFCFF}0.00 & \cellcolor[HTML]{C7DFB3}0.00   & \cellcolor[HTML]{C8DCB1}-0.30  \\
64 & \cellcolor[HTML]{63BE7B}1.00 & \cellcolor[HTML]{63BE7B}1.00 & \cellcolor[HTML]{63BE7B}1.00 & \cellcolor[HTML]{63BE7B}1.00 & \cellcolor[HTML]{63BE7B}1.00 & \cellcolor[HTML]{FCFCFF}0.00 & \cellcolor[HTML]{C7DFB3}0.00   & \cellcolor[HTML]{C7DFB3}-0.04   \\
 \bottomrule
\end{tabular}
}{}
\vspace{-4mm}
\end{table}



\bibliographystyle{IEEEtran}
{\tiny
\bibliography{refs}
}

\end{document}